# Evolution of the electronic structure of ultrathin MnBi$_2$Te$_4$ Films

**Authors:** Runzhe Xu[1,*], Yunhe Bai[1,*], Jingsong Zhou[1], Jiaheng Li[1], Xu Gu[1], Na Qin[1], Zhongxu Yin[1], Xian Du[1], Qinqin Zhang[1], Wenxuan Zhao[1], Yidian Li[1], Yang Wu[1], Cui Ding[1,5], Lili Wang[1,6], Aiji Liang[2,3], Zhongkai Liu[2,3], Yong Xu[1], Xiao Feng[1,6], Ke He[1,6], Yulin Chen[1,2,3,4,†], and Lexian Yang[1,6,†]

**Affiliations**

[1]State Key Laboratory of Low Dimensional Quantum Physics, Department of Physics, Tsinghua University, Beijing 100084, China.

[2]School of Physical Science and Technology, ShanghaiTech University and CAS-Shanghai Science Research Center, Shanghai 201210, China.

[3]ShanghaiTech Laboratory for Topological Physics, Shanghai 200031, China.

[4]Department of Physics, Clarendon Laboratory, University of Oxford, Parks Road, Oxford OX1 3PU, UK.

[5]Beijing Academy of Quantum Information Sciences, Beijing 100193, China

[6]Frontier Science Center for Quantum Information, Beijing 100084, China.

[*]These authors contribute equally to this work.

**ABSTRACT: Ultrathin films of intrinsic magnetic topological insulator MnBi$_2$Te$_4$ exhibit fascinating quantum properties such as quantum anomalous Hall effect and axion insulator state. In this work, we systematically investigate the evolution of the electronic structure of MnBi$_2$Te$_4$ thin films. With increasing film thickness, the electronic structure changes from an insulator-type with large energy gap to one with in-gap topological surface states, which is, however, still in drastic difference to the bulk material. By surface doping of alkali-metal atoms, a Rashba split band gradually emerges and hybridizes with topological surface states, which not only reconcile the puzzling difference between the electronic structures of the bulk and thin film MnBi$_2$Te$_4$ but also provides an interesting platform to establish Rashba ferromagnet that is attractive for (quantum) anomalous Hall effect. Our results provide important insights into the understanding and engineering of the intriguing quantum properties of MnBi$_2$Te$_4$ thin films.**

The combination of magnetism and topology has delivered many intriguing phenomena in topological quantum materials, including quantum anomalous Hall effect (QAHE) [1-5], topological axion insulator [6-8], magnetic Dirac and Weyl semimetals [9-13], Majorana fermions when superconductivity is involved [14], and giant magneto-optical effects [15, 16]. Particularly, the breaking of time-reversal symmetry by magnetic ordering in topological insulators provides an ideal platform to achieve QAHE for dissipationless transport [1-5]. Recently, $MnBi_2Te_4$ has been identified as the first intrinsic magnetic topological insulator [17-27], exhibiting many extraordinary quantum properties such as QAHE at liquid helium temperature [28], topological phase transition between Chern insulator and axion insulator states [29], and high-Chern-number Chern insulator states [30], which provides both important scientific implications and great application potential.

Nevertheless, there are still many mysteries to be understood in the electronic structure of $MnBi_2Te_4$ [17-20, 31]. First of all, the topological surface states (TSSs) were observed by angle-resolved photoemission spectroscopy (ARPES) only at low photon-energies and show a diminishing energy gap that is immune to magnetic phase transition, in drastic contrast to the theoretical prediction and transport measurements [23-27, 32]. Secondly, the dispersion of the TSSs shows a kink-like structure and is strongly broadened near the Fermi level ($E_F$) [25-27, 32]. Thirdly and also importantly, although the intriguing transport properties have been realized in $MnBi_2Te_4$ thin films, their electronic structure is not adequately investigated by experiments and seems to be distinct from that of the bulk sample [22, 33]. These bewildering facts require a systematic investigation on the electronic structure of $MnBi_2Te_4$ thin films, which is necessary for understanding the interplay between the intrinsic electronics structure, magnetic order, and topology in the system.

In this work, using laser-based ARPES with superb energy and momentum resolutions, we systematically investigate the electronic structure of ultrathin $MnBi_2Te_4$ films as well as its evolution with film thickness and surface decoration. With increasing film thickness, the band structure changes from an insulator-type with large energy gap to one that exhibits in-gap TSSs. We observe drastic difference between the electronic structures of the bulk $MnBi_2Te_4$ and the films even when thicker than 7 SL. Interestingly, with increasing surface doping of alkali-metal atoms, a surface band with large Rashba splitting that possibly originate from quantum confinement effect gradually emerges and strongly hybridizes with the TSSs, which bridges the band structures of the film and bulk $MnBi_2Te_4$. More importantly, the Rashba split band (RSB) on the ferromagnetic surface of $MnBi_2Te_4$ establishes a Rashba ferromagnet which is an attractive platform for anomalous Hall effect (AHE) and QAHE [34, 35]. Our results not only help understand the peculiar electronic structure of $MnBi_2Te_4$ but also shed light on the engineering and application of its intriguing quantum properties.

$MnBi_2Te_4$ exhibits layered structure with van der Waals stacking of Te-Bi-Te-Mn-Te-Bi-Te SLs, forming a rhombohedral lattice (space group of $R\bar{3}m$), which can be viewed as inserting an extra Mn-Te layer into the quintuple layer of $Bi_2Te_3$ [31, 36, 37] (Figure 1a). ARPES measurements can be performed either on the bulk crystals after cleavage or on the films synthesized by molecular beam epitaxy (MBE), as schematically shown in Figure 1b and 1c respectively. Compared with bulk crystals grown by the solid-reaction method, MBE can accurately control the chemical composition and the doping level of the films. Figure 1d and 1e compare the electronic structures of 5 SL film and bulk MnBi$_2$Te$_4$. In the bulk sample, we observe sharp TSSs forming a Dirac cone near 270 meV below $E_F$. In contrast to the theoretical calculations, the TSSs show a diminishing band gap at the Dirac point, in agreement with previous reports [25-27]. Peculiarly, the dispersion of the TSSs shows a kink-like structure near -100 meV, above which the TSSs are strongly broadened. In the 5SL film, the TSSs likewise show a diminishing gap. But the TSSs merge into bulk conduction band (BCB) without kink-like structure (Figure 1e). Besides, the doping level of the film is also different from the bulk sample, as manifested by the location of the Dirac point. Except for the drastic difference of the TSSs, the bulk band structure, including the dispersion and the bulk band gap, is quite similar in the film and bulk sample, suggesting that the surface situation and doping level are crucial for the dispersion of the TSSs. Interestingly, surface doping of alkali metals, by applying an effective electric field on the sample surface (Figure 1f), can effectively tune both the surface situation and doping level, which has been widely applied to modulate the electronic structure of different materials [38-42]. We will use this method to investigate the difference between the electronic structures of the film and bulk MnBi$_2$Te$_4$.

The Fermi surface (FS) of MnBi$_2$Te$_4$ films consists of an electron pocket around $\bar{\Gamma}$ [Figure 2a(i-v)], which is clearly different from that of the bulk material showing two electron pockets together with a distribution of blurred spectral weight at $\bar{\Gamma}$ [Figure 2a(vi)]. We observe drastic changes of the band structure along $\bar{\Gamma}\bar{M}$ with film thickness. For the 1 SL film, there is an energy gap larger than 300 meV between the valence band and conduction band. As the film thickness increases, the 2 SL film shows a Dirac-cone like dispersion with a band gap of about 60 meV between the conduction and valence bands (see Figure S3). Subsequently, the band gap between the conduction and valence bands increases in the thicker films with the in-gap states emerges and gradually evolves to the surface states in the thick films and the bulk sample. Above 4 SL, no clear change of band structure is observed with further increasing thickness except that the spectral weight of the TSSs is enhanced. According to our *ab-initio* calculations with out-of-plane anti-ferromagnetism (see Figure S2 and Figure S3), the Chern number of odd- and even-number SL films are different. In particular, the Chern number changes from 0 in 1 SL film to 1 in the 3 SL film [21-23]. By contrast, the Chern number of even-number SL films remains 0 with increasing film thickness. Further investigations are required to experimentally verify this interesting topology evolution.

Similar to the bulk sample, the TSSs also show a diminishing band gap. We measure the electronic structure of the films at temperatures from 7 K to 42 K. No clear difference is observed (see Figure S3), suggesting a minor impact of magnetic phase transition on the TSSs, similar to the situation in the bulk sample [17-20, 31]. The drastic difference between the band structures of the films and the bulk sample, however, persist up to 7 SL film. To understand this difference, we modify the sample surface and tune $E_F$ by *in-situ* doping alkali-metal atoms on the sample surface. Figure 3a presents the evolution of the band structure of the 6 SL film with surface potassium (K) doping (data on 5 SL film can be seen in the Figure S4). With slight K doping, the bands rigidly shift towards higher binding energies and the spectral weight of the TSSs gets enhanced with increasing doping level, despite the introduction of surface disorder [Figure 3a(i-iv)]. At 0.07 ML coverage of potassium, an extra band emerges [Figure 3a(iv) and 3b(ii)], which shows an X-shape dispersion at 0.1 ML, mimicking the Rashba split bands (RSBs) on the surface of the topological insulator $Bi_2Se_3$ [43, 44]. Surprisingly, at 0.12 ML coverage, the dispersion of the TSSs shows a kink-like structure near 100 meV [Figure 3a(v)], resembling the band structure of the bulk sample, as better visualized by the second derivative of the spectrum in Figure 3b(iii). With further K doping, the bands first shift towards higher binding energies and then shift backwards. It is worth noting that the RSB and the kink shift synchronously with K doping, suggesting a close relationship between them, which is further supported by the similar spectral weight and broadening of the TSSs above the kink and the RSB. Finally, above 0.24 ML coverage, the RSB together with the kink disappears and the TSSs again terminate at the BCB, while the Dirac point shifts by about 100 meV to higher binding energy. The emergence and disappearance of the RSB, together with the change of the dispersion of the TSSs, will alter the structure of the FS, suggesting a Lifshitz transition induced by surface decoration (Figure S5).

The observed RSB is reminiscent of the Rashba-like band in the bulk $MnBi_2Te_4$ [Figure 1d, 4a(i), and 4b(i)] [26]. To reveal its relationship with the observed RSB in the film, we study the evolution of the band structure of bulk $MnBi_2Te_4$ with surface decoration in Figure 4. The evolution trend is generally similar to that of the films. Upon slight doping, the X-shape RSB becomes sharper and shifts towards higher binding energies. By contrast, Dirac point, the bulk valence band (BVB) and BCB show only a slight shift. Interestingly, the kink in the dispersion of the TSSs also shifts towards higher binding energies and is locked-in with the RSB [Figure 4a(i-vi)]. Again, the TSSs above the kink and the RSB show similar spectral weight and broadening. Moreover, with further doping, the RSB and the kink synchronously shift towards $E_F$ [Figure 4a(vii), 4b(ii), and 4b(iii)] and gradually disappear at heavy doping level [> 0.15 ML, Figure 4a(viii)], clearly suggesting their common origin. Interestingly, the band structure of bulk MnBi$_2$Te$_4$ eventually becomes similar to that of the pristine films, leaving only the BVB, BCB, and TSSs with a diminishing gap [Figure 4a(viii) and (ix)]. The overall evolution of the band structure of the bulk sample also shows a non-monotonic shift with surface K doping. As shown in Figure 4c, the RSB, the BCB, and the BVB first shift towards higher binding energies, then gradually shift back towards $E_F$, similar to the evolution of the band structure of the films in Figure 3.

The observed non-monotonic shift of the band structure, however, is beyond the expectation of the simple electron-donating picture, although it has been observed in different materials [40, 45]. We suggest that the reversed band shift after a critical doping level can be understood by the nucleation of doped alkali atoms, similar to previous observation on Ag/Si(111) surface [46]. This scenario is supported by the observation of micro-crystals or clusters on the heavily doped materials using an atomic force microscope (see Figure S7).

The most important observation in our work is the RSB and its interaction with the TSSs. Based on their similar evolution, we conclude that the RSBs in the film and bulk have the same origin. Using a simple four-band model, our numerical calculation on a MnBi$_2$Te$_4$ slab suggests that the RSB originates from the quantum confinement effect that can be effectively tuned by surface doping (see Figure S8). Similar RSB with large tunability has been observed in Bi$_2$Se$_3$ and attributed to the quantum-confined conduction band states [43, 44]. The sensitivity of the RSB to surface doping suggests its close relationship with sample surface condition. The lock-in evolution of the RSB and the kink of the TSSs suggests their common origin. We argue that the kink is induced by the hybridization between the TSSs and the RSB. As schematically shown in Figure 5a, the RSB is unoccupied in the pristine films. With surface doping, the RSB shifts downward and hybridizes with the TSSs (Figure 5b and 5c), which induces the "kink"-like structure, resembling the TSSs of bulk $MnBi_2Te_4$. With further doping, the RSB shifts backward and finally disappears. The band structure then recovers to that of the pristine film.

The RSB strongly depends on the surface condition but not on the film thickness when thicker than 3 SL. Therefore, we expect it also exists and is occupied in the films that exfoliated from the bulk materials. It will contribute significant carrier densities and play an essential role in the intriguing transport properties such as QAHE of $MnBi_2Te_4$ films. Particularly, in the films with odd number (> 3) of $MnBi_2Te_4$ SLs, the top SL exhibits ferromagnetism while the magnetization of the other layers cancels with each other. Therefore, the surface constructs a ferromagnetic Rashba metallic system, which may provide a possible route to realize QAHE, as proposed by M. Onoda and N. Nagaosa [34, 35].

In conclusion, we systematically investigate the evolution of the electronic structure of $MnBi_2Te_4$ films with film thickness and surface decoration. With increasing film thickness, we reveal a change of the band structure from an insulator-type to one that exhibits in-gap TSSs. With surface doping of alkali metal, we unravel a surface band with large Rashba splitting. The RSB not only contributes significant carriers in the system but also strongly interacts with the TSSs, which provides a convincing explanation of the peculiar kink-like structure in the dispersion of the TSSs observed in the bulk samples. We suggest that the RSB may play an important role in the extraordinary transport properties of $MnBi_2Te_4$ films, which provides new insights into the contradiction between previous ARPES and transport measurements.

**Sample growth.** $MnBi_2Te_4$ films were grown on 0.7% niobium doped $SrTiO_3$ (111) substrates by coevaporation method in an ultrahigh-vacuum molecular beam epitaxy (MBE) system (base pressure is about 1.5×10⁻¹⁰ mbar). The substrate was degassed at 500°C for 2h and annealed at 950°C for 15min before growth. Sharp 1×1 streaks in reflection high energy electron diffraction (RHEED) pattern with Kikuchi lines were obtained. High purity Mn (99.9998%), Bi (99.9999%), Te (99.9999%) were coevaporated with commercial Knudsen cells. The growth of $MnBi_2Te_4$ was carried out under Te-rich condition at the substrate temperature of 270 °C followed by an annealing process for 30 min to further improve sample quality. Sharp streaky RHEED patterns of samples with different thicknesses suggest the high quality of thin films (see Figure S1). The growth rate was about 0.125 SL/min as calibrated by the thickness and coverage of thin films that were checked by an atomic force microscopy (AFM).

The bulk $MnBi_2Te_4$ crystals were grown by solid-reaction method [47]. $Bi_2Te_3$ and MnTe precursors were prepared by reacting high-purity Bi (99.99%, Adamas) and Te (99.999%, Aladdin), and Mn (99.95%, Alfa Aesar) and Te (99.999%, Aladdin), respectively. The mixture of $Bi_2Te_3$ and MnTe with a molar ratio of 1:1 were thoroughly ground into powder and loaded into a vacuum-sealed silica ampoule. The ampoule was heated to 1173 K to homogenize the $Bi_2Te_3$ and MnTe precursors, followed by a slow cool-down process to 864 K at a rate of 2K per hour. The crystal growth then occurred in the long-term annealing at 864 K that afforded mm-sized and shiny crystalline flakes of $MnBi_2Te_4$. The crystals were examined on a PANalytical Empyrean diffractometer with Cu Kα radiation. High-quality single crystals were selected and stored for the experiment.

**Laser-based ARPES.** Laser-based ARPES measurements were performed using DA30 analyzer and vacuum ultraviolet 7 eV lasers in Tsinghua University, China. The overall energy and angular resolutions were set to better than 5 meV and 0.2°, respectively. The bulk samples were cleaved *in situ* and measured under ultra-high vacuum below 6.5 × 10$^{-11}$ mbar. The films were transferred to ARPES system via a vacuum suitcase with pressure below 1.5 × 10$^{-9}$ mbar. Surface doping was performed *in situ* at 70 K using a SAES alkali-metal source after well outgassing. The current for K source was set to 5.6 A.

**Theoretical calculation.** First-principles calculations were performed in the framework of density functional theory (DFT) by the Vienna *ab-initio* simulation package [48], using the projector augmented wave potential [49], the Perdew-Burke-Ernzerhof exchange-correlation functional [50], and plane wave basis with 350 eV energy cutoff. The GGA+$U$ method [51] was adopted to describe localized 3$d$ orbitals of Mn atoms, using $U$ = 4eV as previously tested [23]. The DFT-D3 method was applied to introduce van der Waals corrections [52]. The spin-orbit coupling was included in all the calculations. The structural relaxation was performed with a force criterion of 0.01 eV/Å. The Γ-centered 9 × 1 × 1 $k$-mesh was adopted for self-consistent calculations of thin films.

## ASSOCIATED CONTENT

## Supporting Information

The Supporting Information is available free of charge at [Nano Letters website].

More information on sample growth, characterization, and evolution of the electronic structure with temperature and surface doping.

## AUTHOR INFORMATION

## Corresponding Authors

‡Email: LXY: lxyang@tsinghua.edu.cn; YLC: yulin.chen@physics.ox.ac.uk

## Author contributions

L.X.Y. and Y.L.C. conceived the experiments. R.Z.X. carried out ARPES measurements with the assistance of J.S.Z., X.G., N.Q., Z.X.Y., X.D., Q.Q.Z., W.X.Z., Y.D.L., A.J.L. and Z.K.L.; R.Z.X. performed the data analysis on the ARPES results; Y.H.B. performed the growth of high quality $MnBi_2Te_4$ thin films supervised by K.H. and X.F.; Y.W synthesized the single crystal $MnBi_2Te_4$ for measurement; J.H.L. performed ab initio calculations with assistance of Y.X.; C.D. performed the STM experiments guided by L.L.W.; R.Z.X. wrote the first draft of the paper. L.X.Y., Y.L.C. and Z.K.L. contributed to the revision of the manuscript. All authors contributed to the scientific planning and discussion.

## Acknowledgments

This work was supported by the National Natural Science Foundation of China (Grants No. 11774190, No. 11427903, No. 11634009, No. 21975140, and No. 51991343), the National Key R&D program of China (Grants No. 2017YFA0304600, No. 2017YFA0305400, and No. 2017YFA0402900), and EPSRC Platform Grant (Grant No. EP/M020517/1). L. X. Y. acknowledges the support from Tsinghua University Initiative Scientific Research Program.

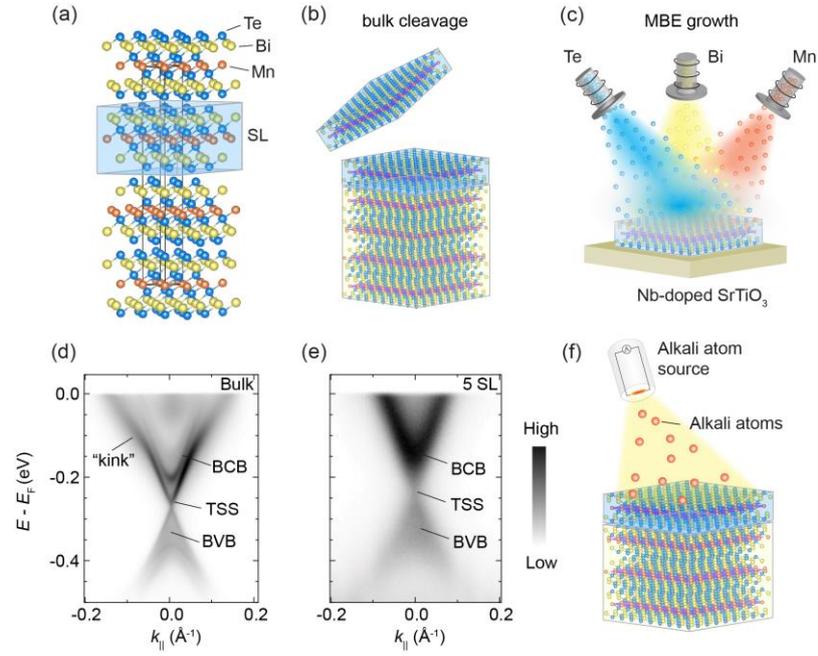

**Figure 1.** Schematic crystal and electronic structure of MnBi$_2$Te$_4$ thin film and bulk crystal. (a) Crystal structure of MnBi$_2$Te$_4$ showing the stacking van der Waals septuple layers. (b) Cleavage of the bulk material to obtain clean surface. (c) Growth of MnBi$_2$Te$_4$ films using molecular beam epitaxy (MBE). (d, e) Band structures of MnBi$_2$Te$_4$ bulk sample (d) and 5 SL film (e) along $\bar{\Gamma}\bar{M}$. BCB: bulk conduction band. BVB: bulk valence band. TSS: topological surface state. (f) Schematics of the surface doping of alkali-metal atoms, which applies an effective voltage on the sample surface. Data in (d) and (e) were measured using 7 eV lasers at 19 K and 7 K respectively.

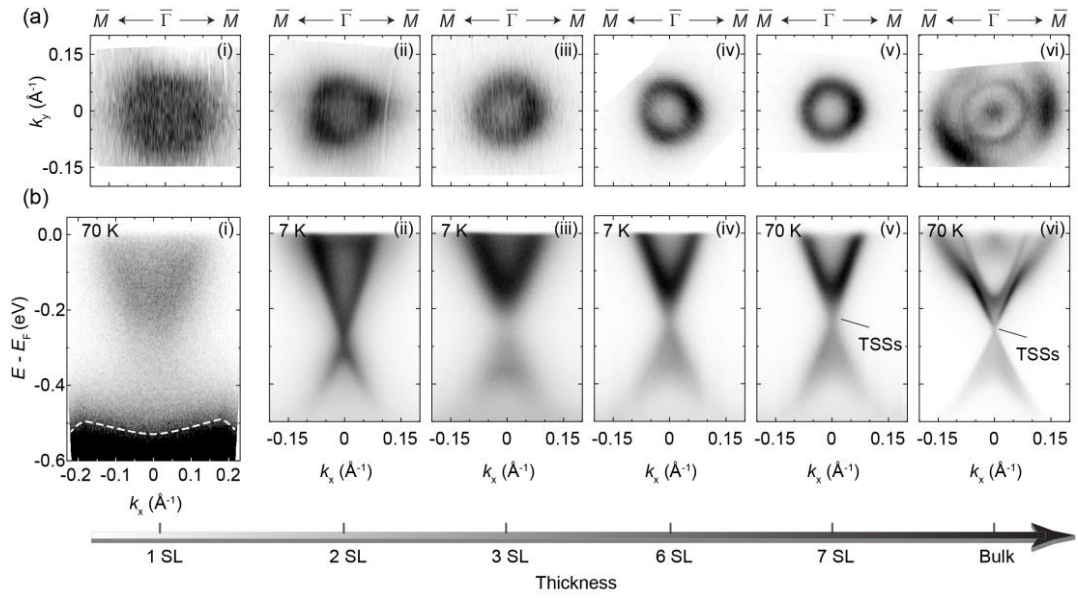

**Figure 2.** Evolution of the electronic structure of MnBi$_2$Te$_4$ films with film thickness. (a) Fermi surfaces of MnBi$_2$Te$_4$ films and bulk crystal. (b) ARPES intensity maps along $\bar{\Gamma}\bar{M}$. The white dashed line is the guide of eyes for the M-shape valence band in 1 SL film. Data were collected using 7 eV laser.

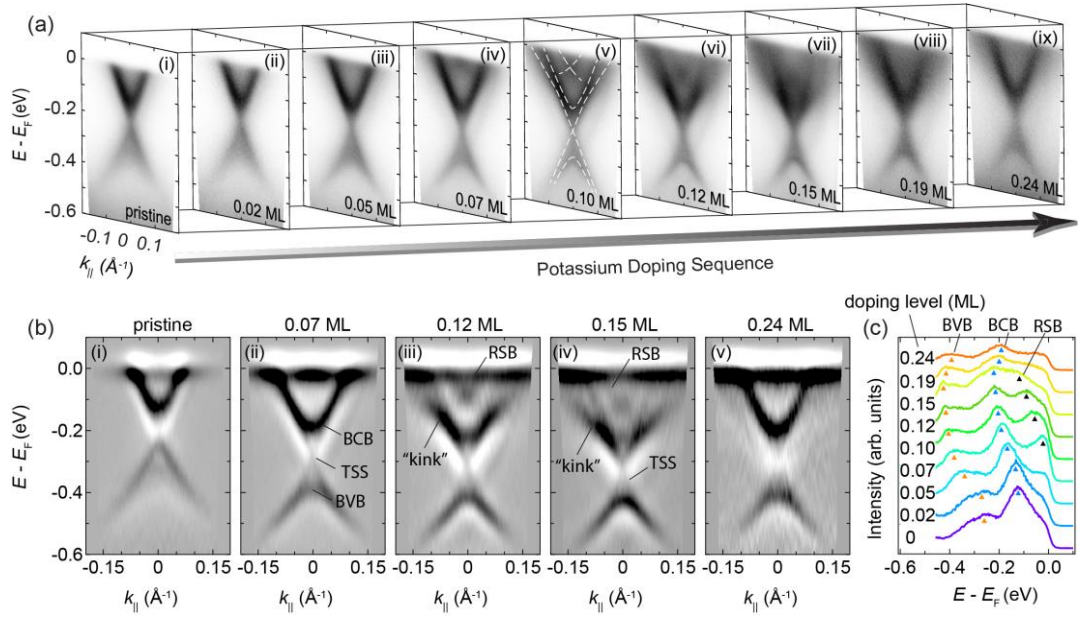

**Figure 3.** Evolution of the band structure of 6 SL thin film with surface K doping. (a) Evolution of the band dispersion along $\bar{\Gamma}\bar{M}$ with K doping level. The white dashed lines in (a)(v) are guide to the eyes for the band dispersions. (b) Second derivative of the spectra at selected doping levels. (c) Evolution of the energy distribution curves (EDCs) at $\bar{\Gamma}$ with the doping level. Data were collected at 70 K.

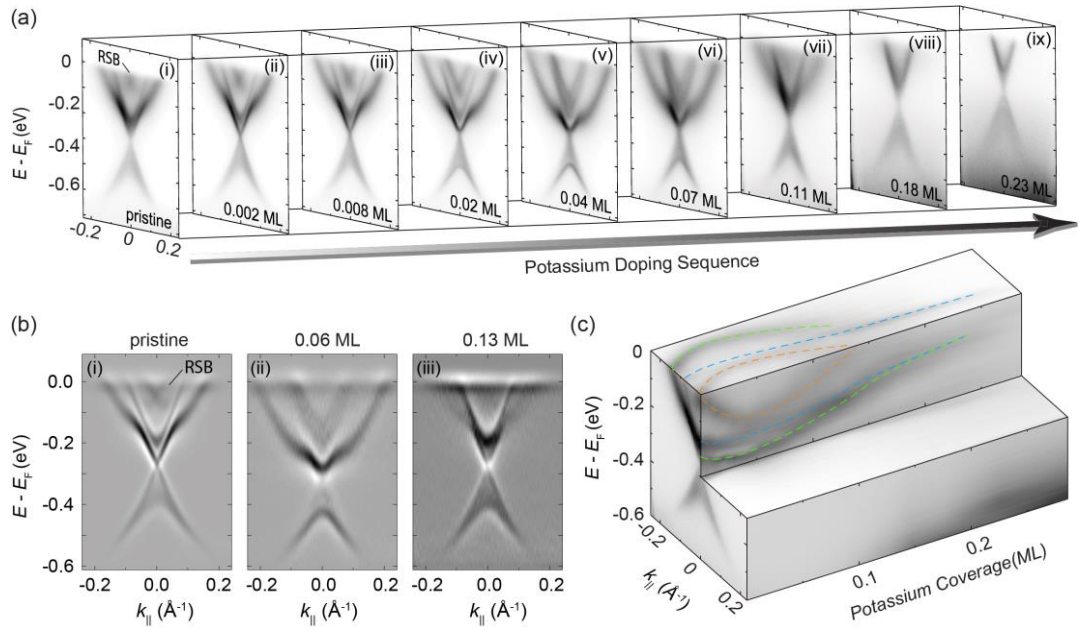

**Figure 4.** Evolution of the band structure of bulk MnBi$_2$Te$_4$ with surface K doping. (a) Doping evolution of the band dispersion along $\bar{\Gamma}\bar{M}$. (b) Second derivative of the spectra at selected doping levels. (c) 3D plot of the non-monotonic shift of the energy bands. Data were collected at 70 K.

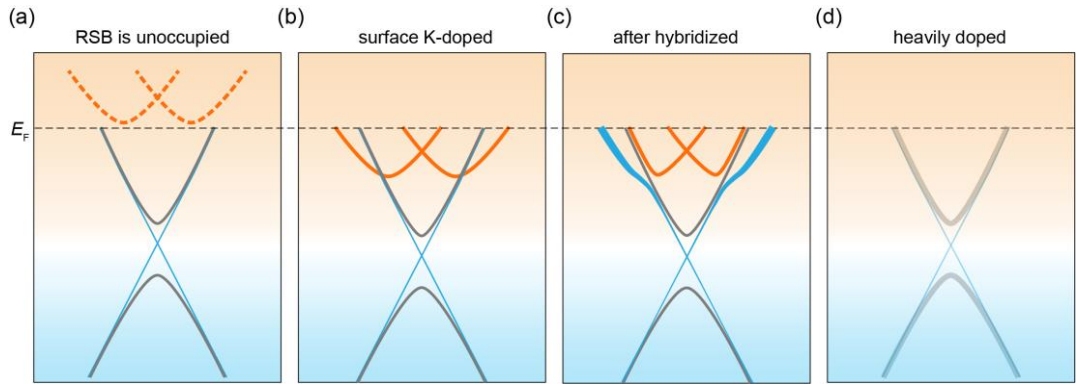

**Figure 5.** Schematic illustration of the evolution of the band structure of MnBi$_2$Te$_4$ film with surface decoration and the origin of the "kink" in the dispersion of TSSs. (a) The Rashba-split band (RSB, orange dashed lines) is unoccupied in the pristine films. (b) The RSB is sensitive to the surface doping and shifts to higher binding energies. (c) The RSB hybridizes with the topological surface band (TSB), which induces the "kink"-like structure that is commonly observed in bulk MnBi$_2$Te$_4$. (d) With heavy doping, the RSB is suppressed and the band structure recovers to that of the pristine films.